\documentclass[prl,aps,twocolumn,floatfix,floats]{revtex4}
\usepackage{epsfig}
\usepackage{amsmath}
\usepackage{amssymb}
\usepackage{mathrsfs}
\usepackage{color}
\usepackage{subfigure}
\usepackage{url}
\usepackage{enumerate}
\usepackage{braket}
\usepackage{comment}
\newcommand{\be}{\begin{equation}}
\newcommand{\ee}{\end{equation}}
\newcommand{\bea}{\begin{eqnarray}}
\newcommand{\eea}{\end{eqnarray}}
\newcommand{\ba}{\begin{aligned}}
\newcommand{\ea}{\end{aligned}}

\def\nn{\nonumber\\}
\def\fr#1{(\ref{#1})}

\newcommand\veps{\varepsilon}
\def\ve{\varepsilon}

\def\XXint#1#2#3{{\setbox0=\hbox{$#1{#2#3}{\int}$}
     \vcenter{\hbox{$#2#3$}}\kern-.5\wd0}}

\begin{document}

\title{Dynamical Correlations after a Quantum Quench
}%

\author{Fabian H.L. Essler}
\author{Stefano Evangelisti}
\author{Maurizio Fagotti}
\affiliation{\mbox{The Rudolf Peierls Centre for Theoretical Physics,
    Oxford University, Oxford, OX1 3NP, United Kingdom}}
\begin{abstract}
We consider dynamic (non equal time) correlation functions of local
observables after a quantum quench. We show that in the absence of
long-range interactions in the final Hamiltonian, the dynamics is
determined by the same ensemble that describes static (equal time)
correlations. For many integrable models static correlation functions
of local observables after a quantum quench relax to stationary values,
which are described by a \emph{generalized Gibbs ensemble} (GGE). The
same GGE then determines dynamic correlation functions and the basic
form of the fluctuation dissipation theorem holds, although the
absorption and emission spectra are not simply related as in the
thermal case. For quenches in the transverse field Ising chain (TFIC)
we derive explicit expressions for the time evolution of dynamic order
parameter correlators after a quench.
\end{abstract}

\maketitle

\paragraph{Introduction.} %
By virtue of their weak coupling to the environment ultra-cold atomic
gases provide ideal testing grounds for studying nonequilibrium
dynamics in isolated many-particle quantum systems. Recent
experiments~\cite{uc,kww-06,tc-07,tetal-11,cetal-12,getal-11}
have observed essentially unitary time evolution on long
time scales. This has stimulated much theoretical research on
fundamental questions such as whether observables generically relax to
time independent values, and if they do, what principles determine their
stationary properties. Relaxational behaviour at first may appear
surprising, because unitary time evolution maintains the system in a
pure state at all times. However, it can be understood intuitively as
a property of a given finite subsystem in the thermodynamic limit,
with the role of the bath being played by the rest of the system.

Dimensionality and conservation laws strongly affect the
out-of-equilibrium dynamics. Ground breaking experiments by Kinoshita, 
Wenger and Weiss~\cite{kww-06} on trapped ${}^{87}{\rm Rb}$ atoms
established that three  dimensional condensates ``thermalize''
rapidly, i.e. relax quickly to a stationary state characterized by an
effective temperature, whereas the relaxation of quasi one-dimensional
systems is slow and towards an unusual non-thermal distribution. 
This difference has been attributed to the presence of
approximate conservation laws in the quasi-1D case, which are argued
to constrain the dynamics. The findings of Ref.~\cite{kww-06}
sparked a tremendous theoretical effort aimed at clarifying the
effects of quantum integrability on the non-equilibrium evolution in
many-particle quantum systems, see
e.g. Refs~\cite{rev,gg,rdo-08,cc-07,caz-06,bs-08,rsms-08,mwnm-09,fm-10,bkl-10,bhc-10,gce-10,rf-11,cic-11,mc-12,ck-12}
and references therein. A widely held view, that has emerged from
these studies, is that the reduced density matrix of any
\emph{subsystem} (which determines correlation functions of all local
observables within the subsystem) is described in terms of either an
effective thermal (Gibbs) distribution or a so-called generalized
Gibbs ensemble (GGE)~\cite{gg}. The former is believed to represent
the generic case, while substantial evidence suggests that the latter
arises for integrable models. 

Theoretical research so far has focussed on static properties in the
stationary state. A question of both great experimental relevance and
theoretical interest is what characterizes the \emph{dynamical}
properties at late times after a quench. These can be accessed by
experimental probes at finite energies, such as photoemission
spectroscopy~\cite{ARPES}. In the first part of this letter we prove
quite generally, that dynamical correlations of local
operators acting within a given subsystem in the stationary state
after a quantum quench are determined by the same distribution
function as static correlations. In particular this
means that whenever the GGE describes static correlations in the 
stationary state, it also applies to the dynamics. 

\paragraph{Stationary State Dynamics after a Quantum Quench.}
We consider the following quench protocol. The system is prepared in
the ground state $|\Psi_0\rangle$ of a lattice Hamiltonian $H(h_0)$ with local
interactions, where $h_0$ is a system parameter such as a magnetic
field. At time $t=0$ we suddenly change $h_0$ to $h$ and the system
time evolves unitarily with Hamiltonian $H(h)$ thereafter. We are
interested in expectation values of the form ($t_1,\dots,t_n>0$)
\be
\langle \Psi_0(t)|{\cal O}_1(t_1)\ldots{\cal
  O}_n(t_n)|\Psi_0(t)\rangle\, ,
\ee
where ${\cal O}_j$ are local observables.
We wish to demonstrate the following. If the stationary state of a
quantum many-body system after a quantum quench is described by a
density matrix $\rho_{\rm stat}$ such that for observables ${\cal
  O}_j$ acting only within a subsystem $S$ one has
\be
\lim_{t\to\infty}\langle\Psi_0(t)|
{\cal  O}_1\ldots{\cal O}_n|\Psi_0(t)\rangle
={\rm Tr}\big(\rho_{\rm stat}{\cal
  O}_1\ldots{\cal O}_n\big)\, ,
\label{static}
\ee
then dynamical correlations are described by the same density matrix,
i.e. for $t_1,\ldots,t_n$ fixed we have
\bea
\lim_{t\to\infty}\langle\Psi_0(t)|
{\cal  O}_1(t_1)\ldots{\cal O}_n(t_n)|\Psi_0(t)\rangle\nn
={\rm Tr}\big(\rho_{\rm stat}{\cal
  O}_1(t_1)\ldots{\cal O}_n(t_n)\big)\, .
\label{dynamic}
\eea
The proof of this statement is based on the Lieb-Robinson
bound~\cite{LR:1972} and more specifically the following 
theorem by Bravyi, Hastings and Verstraete~\cite{BHV:2006}: 
let $O_A$ be an operator that differs from the identity only
within a local region $A$. Now define the projection of the (non-local)
operator $O_A(t)$ to the subsystem $S\supset A$ by
\be
O^{(S)}_A(t)\equiv \frac{\mathrm{tr}_{\bar{S}}[
    O_A(t)]\otimes \mathrm I_{\bar{S}}}{\mathrm{tr}_{\bar{S}}[\mathrm I_{\bar{S}}]}\, ,
\ee
where $\bar{S}$ is the complement of $S$. If the time evolution is
induced by a short-range lattice Hamiltonian, then
\be\label{eq:upbound}
\lVert O_A(t)- O_A^{(S)}(t)\rVert\leq c|A|
e^{-\frac{d-v|t|}{\xi}}\, ,
\ee
where $\lVert.\rVert$ is the operator norm, $v$ is the maximal velocity at
which information propagates~\cite{LR:1972}, $d$ is the (smallest)
distance between $\bar{S}$ and A, $|A|$ is the number of vertices in
set $A$, and $\xi$, $c$ positive constants. Assuming the operator
$\mathcal O_2$ to be bounded, $||\mathcal O_2||\leq \kappa$, we therefore have
\bea
\label{eq:ineq0}
&&|\langle \delta\mathcal O_1(t_1)
\mathcal O_2(t_2)\rangle_t|\leq
\lVert\delta\mathcal O_1(t_1)
\mathcal O_2(t_2)\rVert\nn
&&\ \leq\lVert\delta\mathcal O_1(t_1)\rVert\
\lVert \mathcal O_2(t_2)\rVert
\leq c_1|A_1| \kappa\ e^{-\frac{d_1-v|t_1|}{\xi}}\, ,
\eea
where $\braket{.}_t$ denotes expectation value with respect to
$\ket{\Psi_0(t)}$ and $\delta \mathcal O_1(t)=\mathcal O_1(t)-\mathcal
O_1^S(t)$. 
The first inequality holds because the operator norm is an upper bound
for the expectation value on any state, while in the last step
we used \fr{eq:upbound}. Eqn~\eqref{eq:ineq0} implies that
\be
\label{eq:ineq01}
\langle\prod_{j=1}^2 \mathcal O_j(t_j)\rangle_t=
\langle\mathcal O_1^S(t_1)\mathcal O_2(t_2)\rangle_t+
a_1(t_1,t_2,t)e^{-\frac{d_1-v|t_1|}{\xi}}\, ,
\ee
where $a_1(t_1,t_2,t)$ is a bounded function. By repeating the steps
leading to \fr{eq:ineq01} for the operators $\mathcal O_2(t_2)$ we arrive at
$\braket{\mathcal O_1(t_1)\mathcal O_2(t_2)}_t=\braket{\mathcal
  O_1^S(t_1)\mathcal O_2^S(t_2)}_t
+\sum_{i=1}^2 a_i(t_1,t_2,t)\exp\big(-\frac{d_i-v|t_i|}{\xi}\big)$,
where $a_2(t_1,t_2,t)$ is another bounded function. We may now use
the assumption~\eqref{static} for the expectation value on the right
hand side since all operators act within subsystem $S$
\begin{multline}
\lim_{t\rightarrow\infty}\braket{\mathcal O_1(t_1)\mathcal O_2(t_2)}_t=\mathrm{Tr}(\rho_{\rm stat}\mathcal O_1^S(t_1)\mathcal O_2^S(t_2) )\\
+\sum_{i=1}^2 a_i(t_1,t_2)e^{-\frac{d_i-v|t_i|}{\xi}}\, ,
\label{eq:ineq1a}
\end{multline}
where $\lim_{t\to\infty}a_i(t_1,t_2,t)=a_i(t_1,t_2)$ is assumed to
exists for simplicity\endnote{We can drop this assumption and bound
the sum on the r.h.s. in a $t$-independent way instead. The
corresponding contributions then vanish in the limit of an infinitely
large subsystem.}.
The chain of inequalities~\eqref{eq:ineq0} also holds
for the average with respect to the density matrix
$\rho_{\rm  stat}$, i.e.
\begin{multline}\label{eq:ineq2}
\mathrm{Tr}\big(\rho_{\rm stat}\mathcal O_1(t_1)\mathcal O_2(t_2)\big)=
\mathrm{Tr}\big(\rho_{\rm stat}\mathcal O_1^S(t_1)\mathcal O_2^S(t_2)\big)+\\
\sum_{i=1}^2 b_i(t_1,t_2)e^{-\frac{d_i-v|t_i|}{\xi}}\, ,
\end{multline}
where $b_i(t_1,t_2)$ are bounded functions of $t_{1,2}$. Finally,
combining \fr{eq:ineq1a} and \fr{eq:ineq2} and then taking the size of
the subsystem $S$ to be infinite we obtain \fr{dynamic} in the case
$n=2$. The generalization to arbitrary $n$ is straightforward.

\paragraph{Generalized Gibbs ensemble.}
We now concentrate on a quantum quench in an integrable model
in one dimension with Hamiltonian $H(h)\equiv I_1$ and
\emph{local} conservation laws $I_{n\geq 1}$, i.e. $[I_m,I_n]=0$.
The full (reduced) density matrix of the system (of a subsystem $A$)
at time $t$ after the quench is
\be
\rho(t)=|\Psi_0(t)\rangle\langle\Psi_0(t)|\, ,\quad
\rho_A(t)={\rm Tr}_{\bar{A}}\big(\rho(t)\big)\, ,
\ee
where $\bar{A}$ is the complement of $A$. It is widely believed, and was
shown for quenches of the transverse field in the TFIC in
Refs~\cite{CEF:2011,CEF2:2012}, that
\be
\lim_{t\to\infty}\rho_A(t)={\rm Tr}_{\bar{A}}\big(\rho_{\rm GGE}\big)\, ,
\label{GGE_static}
\ee
where
\be
\rho_{\rm GGE}=\frac{1}{Z_{\rm GGE}}e^{-\sum_m \lambda_m I_m}\, ,
\label{GGE}
\ee
is the density matrix of the GGE and ${Z_{\rm GGE}}$ 
ensures the normalization $\rm{tr}\big(\rho_{\rm GGE}\big)=1$.
Eqn~\fr{GGE_static} establishes that all \emph{local, equal time}
correlation functions of a given subsystem in the stationary 
state are determined by the GGE~\fr{GGE}. Applying our result~\fr{dynamic}
 to the case at hand, we conclude that dynamic correlation
functions are also given by the GGE, i.e.
\bea
\lim_{t\to\infty}\langle\Psi_0(t)|
{\cal  O}_1(t_1)\ldots{\cal O}_n(t_n)|\Psi_0(t)\rangle\nn
={\rm Tr}\big(\rho_{\rm GGE}{\cal
  O}_1(t_1)\ldots{\cal O}_n(t_n)\big)\, .
\label{dynamicsGGE}
\eea

\paragraph{Fluctuation Dissipation Relation (FDR).}
A key question regarding dynamical properties in the stationary
state after a quench is whether a FDR holds~\cite{FCG:2011}. Given the
result~\fr{dynamicsGGE}, we can answer this question for cases where
the stationary state is either described by a thermal distribution
with effective temperature $T_{\rm eff}$ or by a GGE. In the former
case, the standard thermal FDR with temperature $T_{\rm eff}$ applies.
The GGE case is more involved and we turn to it next. The linear
response function of observables $A_j$ and $B_l$ acting on sites $j$
and $l$ of a translationally invariant lattice of $L$ sites is 
\bea
\chi_{AB}(\omega,{\bf q})=-\frac{i}{L}\sum_{j,l}\int_0^\infty d\tau
e^{i\omega \tau-i{\bf q}({\bf r}_j-{\bf r}_l)}\nn
\times\ \mathrm{tr}[\rho_{\rm GGE}[A_j(\tau),B_l]]\, .
\eea
On the other hand, the spectral function of the same two
observables in the stationary state is given by 
\bea
S_{A B}(\omega,{\bf q})=\frac{1}{L}\sum_{j,l}
\int_{-\infty}^\infty\frac{\mathrm d \tau}{2\pi}
  e^{i\omega \tau-i{\bf q}\cdot({\bf r}_l-{\bf r}_j)}\nn
\times\ \mathrm{tr}[\rho_{\rm GGE}A_{l}(\tau) B_j]\, .
\eea
Using a Lehmann representation in terms of Hamiltonian eigenstates it
is straightforward to show that
\be
-\frac{1}{\pi}\mathrm{Im}\ \chi_{AB}(\omega,{\bf q})=
S_{AB}(\omega,{\bf q})-S_{BA}(-\omega,-{\bf q})\, ,
\ee
i.e. the basic form of the FDR holds. However, as was already noted in
Ref.~\cite{FCG:2011} for the TFIC, unlike in the thermal
(Gibbs) case, the negative frequency part $S_{BA}(-\omega,-{\bf q})$ is
not related to the positive frequency part by a simple relation of the
form $S_{AB}(-\omega,-{\bf q})=f(\omega)S_{BA}(\omega,{\bf q})$,
where $f(\omega)$ is independent of $A$ and $B$.

\paragraph{Transverse Field Ising Chain.}
We now focus on the dynamics after a quantum quench in a particular
example, the TFIC described by the Hamiltonian
\be
H(h)=-J\sum_{j=1}^L\Bigl[\sigma_j^x\sigma_{j+1}^x+h\sigma_j^z\Bigr]\, ,
\label{Hamiltonian}
\ee
where $\sigma_j^\alpha$ are the Pauli matrices at site $j$, $J>0$ and we
impose periodic boundary conditions $\sigma_{L+1}^\alpha=\sigma^\alpha_1$. 
The model~\fr{Hamiltonian} is a crucial paradigm of quantum critical
behaviour and quantum phase transitions~\cite{sachdev}. 
At zero temperature and in the thermodynamic limit it exhibits
ferromagnetic ($h < 1$) and paramagnetic ($h > 1$) phases, separated
by a quantum critical point at $h_c=1$. 
For $h<1$ and $L\rightarrow\infty$ there are two degenerate ground
states. Spontaneous symmetry breaking selects a unique ground state,
in which spins align along the $x$-direction. On the other hand, for
magnetic fields $h>1$ the ground state is non-degenerate and, as the
magnetic field $h$ is increased, spins align more and more along the
$z$-direction.  The order parameter for the quantum phase transition
is the ground state expectation value $\braket{\sigma^x_j}$.
We note that the model~\fr{Hamiltonian} is (approximately) realized
in systems of cold Rb atoms confined in an optical lattice~\cite{ising-ca}.

Two point dynamical correlation functions are of particular importance
due to their relationships to response functions measured in
photoemission and scattering experiments. The two-point function of
transverse spins
$\langle\Psi_0(t)|\sigma^z_{j+\ell}(\tau_1)\sigma^z_j(\tau_2)|\Psi_0(t)\rangle$
in the TFIC can be calculated by elementary means~\cite{mc}
as it is local in terms of Jordan-Wigner fermions. Our goal is to
determine the dynamical order-parameter two-point function 
\be
\rho^{xx}(\ell,t+\tau_1,t+\tau_2)=\langle\Psi_0(t)|\sigma^x_{1+\ell}(\tau_1)
\sigma^x_{1}(\tau_2)|\Psi_0(t)\rangle,
\ee
after quenching the transverse field at time $t=0$ from $h_0$ to $h$
for times $\tau_{1,2}\geq 0$. This can be achieved by employing a
generalization of the form factor methods recently developed in
Ref.~\cite{CEF1:2012} to the non-equal-time case, and augmenting the
results obtained in this way by exploiting the knowledge of exact
limiting behaviours derived in Refs~\cite{CEF1:2012,CEF2:2012}. 
Our approach is outlined in
~\cite{SuppMat}.
For quenches \emph{within the ordered phase} ($h_0,h<1$) we
obtain for large positive $\ell$, $t$ 
\be
\rho^{xx}(\ell,t+\tau,t)\simeq C^x_{\rm FF}(h_0,h)\ R(\ell,\tau,t)\ ,
\label{rhoxx}
\ee
where
\bea
R(\ell,\tau,t)&=&\exp\Big[\int_0^\pi\frac{\mathrm d
    k}{\pi}\log\big(\cos\Delta_k\big)\nn
&\hspace{-1cm}\times&\hspace{-0.65cm}
\min\big\{\max\{\ve^\prime_h(k)\tau,\ell\}, \ve^\prime_h(k)(2t+\tau)\big\}\Big]\, ,\nn
C^x_{\rm FF}(h_0,h)&=&\frac{1-h
  h_0+\sqrt{(1-h^2)(1-h_0^2)}}{2\sqrt{1-h h_0}\sqrt[4]{1-h_0^2}}\, .
\label{R}
\eea
Here \mbox{$\ve_h(k)=2J\sqrt{1+h^2-2h\cos k}$} is the dispersion
relation of elementary
excitations of the Hamiltonian $H(h)$, \mbox{$\cos\Delta_k=
4J^2(1+hh_0-(h+h_0)\cos k)/\ve_h(k)\ve_{h_0}(k)$}
and \mbox{$\ve'_{h}(k)=d\ve_h(k)/dk$}. An important scale in the problem is
given by the ``Fermi-time''
\be
t_F=\frac{\ell}{2v_{\rm max}}\, ,\quad v_{\rm max}={\rm max}_k\ \ve'_h(k)\, ,
\ee
where $v_{\rm max}$ is the maximal propagation velocity of the elementary
excitations of the post-quench Hamiltonian $H(h)$.
We note that the dominant contribution
at large $\ell, t$ \fr{R} has a vanishing imaginary part. This is
similar to the corresponding correlator at finite temperature in
equilibrium~\cite{sachdev}.
In order to assess the accuracy of the asymptotic result~\fr{R} at short and
intermediate times and distances we have computed the correlator~\fr{rhoxx}
numerically on large, open chains by means of a determinant
representation and then extrapolated the results to the thermodynamic
limit. 
\begin{figure}[t]
\includegraphics[width=0.47\textwidth]{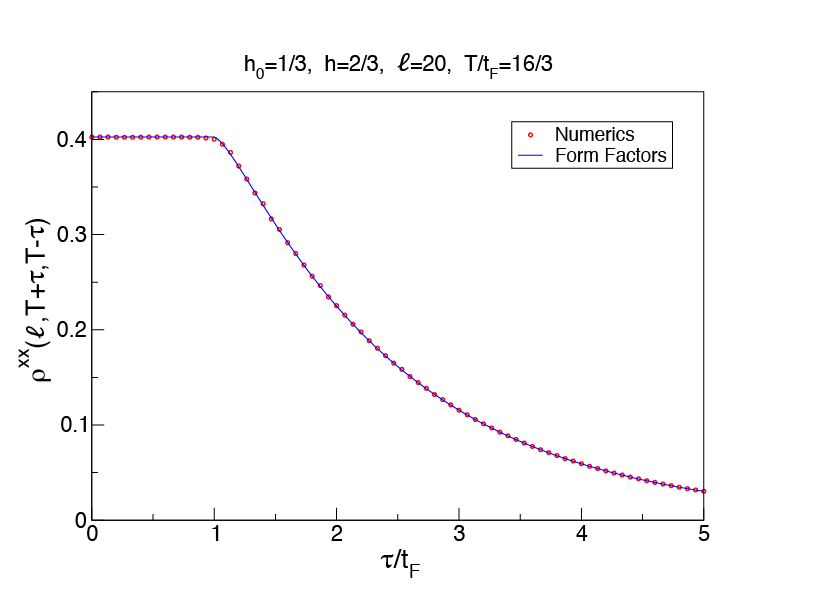}
\caption{Non-equal-time two point function after a quench in the
ordered phase from $h_0=1/3$ to $h=2/3$. The distance and time $T$ are
fixed at $\ell=20$ and $T/t_F=16/3$ respectively.}
\label{fig:ordered}
\end{figure}
A comparison between \fr{rhoxx} and the numerical results for a quench
from $h_0=1/3$ to $h=2/3$ and distance $\ell=20$ is shown in
Fig.~\ref{fig:ordered}. The agreement is clearly excellent. The
qualitative behaviour of $\rho^{xx}(\ell,T+\tau,T-\tau)$ is as
follows: $\tau=0$ corresponds to the known~\cite{CEF1:2012} equal-time
correlator at time $T$ after the quench. The correlator remains
essentially unchanged until $\tau=t_F$ (corresponding to
$\tau_1-\tau_2=2t_F$ in \fr{rhoxx}), where a horizon effect occurs. At
later times $\tau>t_F$ the correlator decays exponentially.

\begin{figure}[t]
\includegraphics[width=0.47\textwidth]{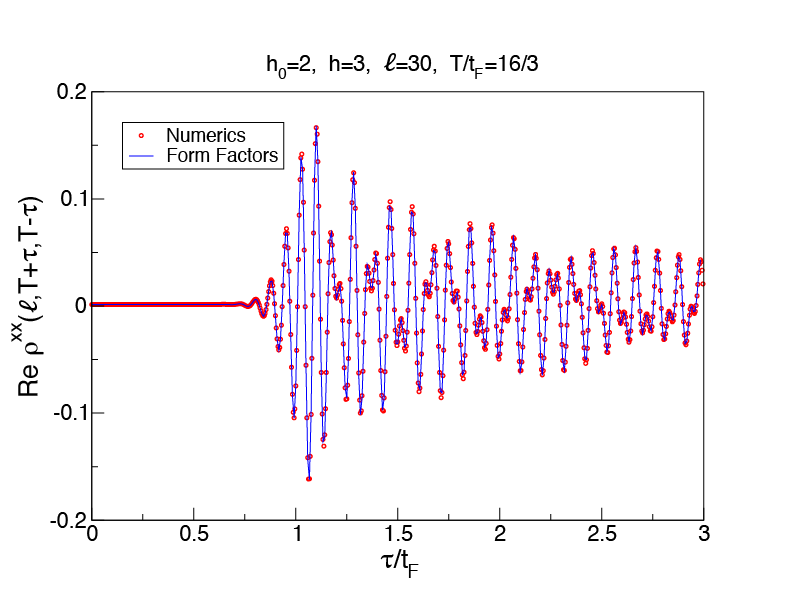}
\caption{Real part of the non-equal-time two point function after a
quench in the disordered phase from $h_0=2$ to $h=3$. The distance and
time $T$ are fixed at $\ell=30$ and $T/t_F=16/3$ respectively. Data
points are numerical results (see the text for details) and the solid
line is eqn~\fr{rhoxxdiso}.}
\label{fig:1}
\end{figure}

\begin{figure}[t]
\includegraphics[width=0.47\textwidth]{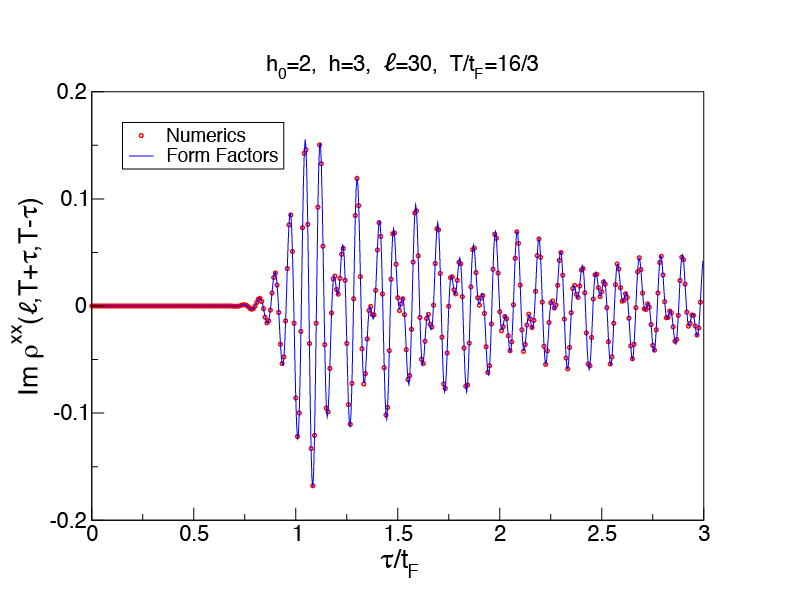}
\caption{Imaginary part of the non-equal-time two point function after a
quench in the disordered phase from $h_0=2$ to $h=3$. The distance and
time $T$ are fixed at $\ell=30$ and $T/t_F=16/3$ respectively. Data
points are numerical results (see the text for details) and the solid
line is eqn~\fr{rhoxxdiso}.}
\label{fig:2}
\end{figure}

For quenches \emph{within the disordered phase} ($h_0,h>1$), we obtain
for $v_{\rm max}(2t+\tau)>\ell$
\be
\rho^{xx}(\ell,t+\tau,t)\simeq
h C^x_{\rm FF}(h_0^{-1},h^{-1}) F(\ell,\tau,t) R(\ell,\tau,t)\, ,
\label{rhoxxdiso}
\ee
where 
$R(\ell,\tau,t)$ and
$C^{x}_{\rm FF}$ are given by \fr{R} 
and 
\begin{multline}
F(\ell,\tau,t)=\int_{-\pi}^\pi \frac{{\mathrm dk}J e^{i\ell k}}{\pi\ve_h(k)}
\Big[e^{-i\ve_k\tau}\\
+2i\tan\big(\frac{\Delta_k}{2}\big)\cos\big(\ve_k(2t+\tau)\big)\mathrm{sgn}\big(\ell-\ve_k^\prime\tau\big)\Big]\, .
\label{F}
\end{multline}
In the complementary regime $v_{\rm max}(2t+\tau)<\ell$ the correlator
is exponentially small and the expressions~(\ref{rhoxxdiso},~\ref{F}) no
longer apply. Outside the ``light-cone'' $v_{\rm max}\tau<\ell$ the
first contribution in \eqref{F} is exponentially small, whereas the
second one decays as a power-law. The result~\fr{rhoxxdiso} is
obtained by a generalization of the form factor~\cite{FFs}
approach developed in
Ref.~\cite{CEF1:2012} and is based on an expansion in the density of
excitations of $H(h)$ in the initial state after the quench. Hence it
is most accurate for quenches where this density is low and breaks
down for quenches from/to the quantum critical point. In
Figs~\ref{fig:1} and \ref{fig:2} we compare the asymptotic result~\fr{rhoxx}
 to numerics obtained in the way described above. The
agreement for the chosen set of parameters ($\ell=30$, $h_0=2$, $h=3$
and $T/t_F=16/3$) is seen to be excellent.
The value of $\rho^{xx}(\ell,T+\tau,T-\tau)$ at $\tau=0$ equals 
the known equal-time correlator at time $T$ after the quench~\cite{CEF1:2012}, 
which is small in the case considered. The correlator remains
largely unchanged up to a horizon at $\tau=t_F$ (corresponding to
$t=t_F/2$ in \fr{rhoxx}), and for times $\tau>t_F$ exhibits an
oscillatory $\tau^{-1/2}$ power-law decay. 
We note that the result~(\ref{rhoxx},~\ref{R}) can be obtained in an
alternative way by generalizing the semiclassical approach of
Ref.~\cite{IR:2011b} (see also \cite{sachdev,rsms-08}) to the
non-equal time case, and then elevating it using exact limiting 
results of Refs~\cite{CEF1:2012,CEF2:2012}. While this method fails
to reproduce the result for quenches in the disordered phase outside
the light-cone, i.e. $v_{\rm max}\tau<\ell$, it provides a 
physical picture. The behaviour is similar to the finite temperature
case \cite{sachdev} and for $h_0,h<1$ can be understood in
terms of classical motion of domain walls. For $h_0,h>1$
(and within the light-cone), quantum fluctuations (associated with the
function $F$ in \fr{F}) give rise to the oscillatory behaviour seen in 
Fig.~\ref{fig:2}, while relaxation occurs at longer scales and
is again driven by classical motion of particles (spin flips) \cite{sachdev}. 
A simple picture emerges when we Fourier transform
$\rho^{xx}(\ell,t+\tau,t)$ at $t\to\infty$ for small quenches. As a
function of $\omega$ for fixed $q$ the resulting ``dynamical
structure factor'' for quenches within the disodered phase 
is dominated by a narrow, asymmetric peak around $\omega=\veps_h(q)$,
while for $h_0,h<1$ we observe a broadening of the
$\delta$-function peak associated with the ferromagnetic order
in the initial state. Both of these are qualitatively similar to the finite-T
equilibrium response \cite{sachdev,finiteT}. 
Having established in the first part of this work that the
$t\to\infty$ limit of $\rho^{xx}(\ell,t+\tau_1,t+\tau_2)$ is described
by the GGE, an important question is how quickly this limiting
behaviour is approached. It follows from (\ref{rhoxx},~\ref{rhoxxdiso}) that
for quenches within the ordered (disordered) phase the limiting value
for fixed $\tau_{1,2}$ and $\ell$ is approached as a $t^{-3}$
($t^{-3/2}$) power law.

\paragraph{Conclusions.}  %
We have considered dynamical correlation functions of local
observables after a quantum quench. We have shown, that dynamical
correlators of local observables in the stationary state are governed
by the same ensemble that describes static correlations. 
For quenches in the TFIC this implies that they
are given by a GGE, for which the basic form of
the fluctuation dissipation theorem holds. We have obtained explicit
expressions for the time evolution of dynamic order parameter
correlators after a quench in the TFIC.  
\acknowledgments
We thank P. Calabrese, J. Cardy, L. Cugliandolo and A. Gambassi for
helpful discussions. This work was supported by the EPSRC under grants
EP/I032487/1 and EP/J014885/1, by an INFN grant (SE) 
and by a Marco Polo fellowship of the University of Bologna (SE).



\appendix
\section{Supplementary Material}
The dynamical two-point functions~(\ref{rhoxx},~\ref{rhoxxdiso}) are
determined by a generalization of the form factor
approach recently developed in Ref.~\cite{CEF1:2012}. The latter is
based on a Lehmann representation of two-point functions in terms of
simultaneous eigenstates of the momentum operator $P$ and the
post-quench Hamiltonian $H(h)$
\bea
H(h)|k_1,\ldots,k_n\rangle_{\tt a}&=&
\Big[\sum_{j=1}^n\varepsilon_h(k_j)\Big]
|k_1,\ldots,k_n\rangle_{\tt a}\, ,\nn
P|k_1,\ldots,k_n\rangle_{\tt a}&=&
\Big[\sum_{j=1}^nk_j\Big]
|k_1,\ldots,k_n\rangle_{\tt a}\, ,
\eea
where ${\tt a}=\rm R,NS$ correspond to periodic/antiperiodic boundary
conditions on the Jordan-Wigner fermions~\cite{CEF1:2012}. For a quench
within the disordered phase the state $|\Psi_0(t)\rangle$ has the
following representation in a large, but finite volume $L$
\be
|\Psi_0(t)\rangle=\frac{|B(t)\rangle_{\rm
    NS}}{\sqrt{{}_{\rm NS}\langle B(t)|B(t)\rangle_{\rm NS}}}\, ,
\ee
where
\begin{multline}
|B(t)\rangle_{\rm
  NS}=\sum_{n=0}^\infty\frac{i^n}{n!}\sum_{0<p_1,\ldots,p_n\in\rm NS} 
\prod_{j=1}^nK(p_j)
e^{-2i\veps_{h}(p_j)t}\\
\times\ |-p_1,p_1,\ldots,
-p_n,p_n\rangle_{\rm NS}\, ,
\end{multline}
\be
K(k)=\frac{\sin(k)\ (h_0-h)}{\frac{\veps_{h_0}(k)
\veps_{h}(k)}{(2J)^{2}}+1+hh_0-(h+h_0)\cos(k)}\, . 
\ee
The function $K(p)$ is related to the quantity $\cos(\Delta_p)$
defined in the main text by $K(p)=\tan(\Delta_p/2)$.
The dynamical order parameter two-point function 
\be
\rho^{xx}(\ell,t+\tau_1,t+\tau_2)=
\frac{{}_{\rm NS}\langle
B(t)|\sigma^x_{m+\ell}(\tau_1)\sigma^x_m(\tau_2)|B(t)\rangle_{\rm NS}}
{{}_{\rm NS}\langle B|B\rangle_{\rm NS}}
\ee
has the following Lehmann representation
\begin{widetext}
\begin{multline}
{}_{\rm NS}\langle
B(t)|\sigma^x_{\ell+m}(\tau_1)\sigma^x_{m}(\tau_2)|B(t)\rangle_{\rm NS}\!=
\sum_{m,n=0}^\infty\frac{i^{n-m}}{n!m!}
\sum_{\genfrac{}{}{0pt}{}{0<p_1,\ldots,p_n\in {\rm
      NS}}{0<k_1,\ldots,k_m\in{\rm NS}}} 
\left[\prod_{j=1}^nK(p_j)e^{-2i(t+\tau_2)\veps_h(p_j)}
\right]\left[\prod_{l=1}^mK(k_l)e^{2i(t+\tau_1)\veps_h(k_l)}\right]\\
\times\sum_{s=0}^\infty\sum_{q_1,\ldots,q_s\in {\rm R}}
\frac{\prod_{r=1}^se^{i(\tau_2-\tau_1)\veps_h(q_r)+iq_r\ell}}{s!}\
\langle k_m,-k_m,\ldots,k_1-k_1|\sigma^x_m
|q_1,\ldots,q_s\rangle\\
\times \langle q_s,\ldots,q_1|\sigma^x_m
|-p_1,p_1,\ldots, -p_n,p_n\rangle\, ,
\label{lehmannquench2}
\end{multline}
\end{widetext}
\be
{}_{\rm NS}\langle B|B\rangle_{\rm NS}=
\exp\Bigl[\sum_{0<q\in {\rm NS}}\log\bigl(1+K^2(q)\bigr)\Bigr]\, .
\ee
The form factors
\be
\langle k_m,-k_m,\ldots,k_1-k_1|\sigma^x_m |q_1,\ldots,q_s\rangle
\ee
are known exactly~\cite{FFs}, see eqns (109)-(111) of Ref.~\cite{CEF1:2012}. 
The leading behaviour of \fr{lehmannquench2} is
evaluated by considering it as a formal expansion in powers of the
function $K(p)$. As shown in Ref.~\cite{CEF1:2012} this corresponds
to an expansion, where the small parameter is the density
of excitations of the post-quench Hamiltonian $H(h)$ in the initial
state $|\Psi_0(0)\rangle$. We determine the dominant contributions at
large $\ell$, $t$ and $|\tau_1-\tau_2|$ to \fr{lehmannquench2} for a
given order in the formal expansion in powers of $K(p)$, and then
sum these to all orders. The structure of this calculation is similar
to the equal time case ($\tau_1=\tau_2$) considered in
Ref.~\cite{CEF1:2012}, but the details differ substantially 
and will be reported elsewhere. The result of the form
factor calculation for quenches within the disordered phase is
\be
\rho^{xx}(\ell,t+\tau,t)\simeq
2J\sqrt{h}(h^2-1)^\frac{1}{4} F(\ell,\tau,t) R_0(\ell,\tau,t)\, ,
\ee
where the function $F$ is given in \fr{F} and
\begin{multline}
R_0(\ell,\tau,t)=\exp\Big[-2\int_0^\pi\frac{\mathrm d
    k}{\pi}K^2(k)\\
\times \min\Big\{\max\{\ve^\prime_h(k)\tau,\ell\}, 
\ve^\prime_h(k)(2t+\tau)\Big\}\Big]\, .
\end{multline}
We now use that the general structure of the resummation for
$\tau_{1,2}\neq 0$ is the same as for $\tau_{1,2}=0$. This allows us
to go beyond the low-density expansion by exploiting results obtained
in Refs~\cite{CEF1:2012,CEF2:2012} for $\tau_1=\tau_2=0$ by means of
determinant techniques. In this way we arrive at eqn~\fr{rhoxxdiso}.
Quenches within the ordered phase are analyzed in the same way.



\begin{thebibliography}{99}
\bibitem{uc}
M.~Greiner, O.~Mandel, T.~W.~H\"ansch, and I.~Bloch,
Nature {\bf 419} 51 (2002).

\bibitem{kww-06}
T. Kinoshita, T. Wenger,  D. S. Weiss, 
Nature {\bf 440}, 900 (2006).

\bibitem{tc-07}
S. Hofferberth, I. Lesanovsky, B. Fischer, T. Schumm, and J. Schmiedmayer,
Nature {\bf 449}, 324 (2007).

\bibitem{tetal-11}
S. Trotzky Y.-A. Chen, A. Flesch, I. P. McCulloch, U. Schollw\"ock,
J. Eisert, and I. Bloch, 
Nature Physics {\bf 8}, 325 (2012).

\bibitem{cetal-12}
M. Cheneau, P. Barmettler, D. Poletti, M. Endres, P. Schauss, T. Fukuhara, C. Gross, I. Bloch, C. Kollath, and S. Kuhr,
Nature {\bf 481}, 484 (2012).

\bibitem{getal-11}
M. Gring, M. Kuhnert, T. Langen, T. Kitagawa, B. Rauer, M. Schreitl, I. Mazets, D. A. Smith, E. Demler, and J. Schmiedmayer,
arXiv:1112.0013.


\bibitem{rev}
A. Polkovnikov, K. Sengupta, A. Silva, and M. Vengalattore,
Rev. Mod. Phys. {\bf 83}, 863 (2011).

\bibitem{gg} M. Rigol, V. Dunjko, V. Yurovsky,  and M. Olshanii,
Phys. Rev. Lett. {\bf 98}, 050405 (2007).

\bibitem{rdo-08}
M. Rigol, V. Dunjko, and M. Olshanii, Nature {\bf 452}, 854 (2008). 

\bibitem{cc-07} 
P. Calabrese and  J. Cardy, J. Stat. Mech. (2007) P06008.

\bibitem{caz-06}
A. Iucci and M. A. Cazalilla, Phys. Rev. A, 80, 063619 (2009).

\bibitem{bs-08}
T. Barthel and U. Schollw\"ock, Phys. Rev. Lett. {\bf 100}, 100601 (2008).

\bibitem{rsms-08}
D. Rossini, A. Silva, G. Mussardo, and G. Santoro, 
Phys. Rev. Lett. {\bf 102}, 127204 (2009); 
D. Rossini, S. Suzuki, G. Mussardo, G. E. Santoro, and A. Silva,
Phys. Rev. B {\bf 82}, 144302 (2010).

\bibitem{fm-10}
D. Fioretto and G. Mussardo, New J. Phys. {\bf 12}, 055015 (2010). 

\bibitem{bkl-10} 
G. Biroli, C. Kollath, and A.M. L\"auchli, Phys. Rev. Lett. {\bf 105}, 250401 (2010).

\bibitem{bhc-10}
M. C. Ba\~nuls, J. I. Cirac, and M. B. Hastings,
Phys. Rev. Lett. {\bf 106}, 050405 (2011).

\bibitem{gce-10}
C. Gogolin, M. P. M\"uller, and J. Eisert,
Phys. Rev. Lett. {\bf 106}, 040401 (2011).

\bibitem{rf-11}
M. Rigol and M. Fitzpatrick,
Phys. Rev. A {\bf 84}, 033640 (2011).

\bibitem{cic-11}
M. A. Cazalilla, A. Iucci, and M.-C. Chung, 
Phys. Rev. E {\bf 85}, 011133 (2012).

\bibitem{mc-12}
J. Mossel and J.-S. Caux, arXiv:1201.1885.

\bibitem{ck-12}
J.-S. Caux and R. M. Konik, arXiv:1203.0901.

\bibitem{mwnm-09}
S. R. Manmana, S. Wessel, R. M. Noack, and A. Muramatsu,
Phys. Rev. B {\bf 79}, 155104 (2009). 

\bibitem{ARPES}
B. Fr\"ohlich, M. Feld, E. Vogt, M. Koschorreck, W. Zwerger and M. K\"ohl,
Phys. Rev. Lett. 106, 105301 (2011).

\bibitem{LR:1972} E. H. Lieb and D. W. Robinson, Commun. Math. Phys. {\bf 28}, 251 (1972)

\bibitem{BHV:2006} S. Bravyi, M. B. Hastings, and F. Verstraete, Phys. Rev. Lett. {\bf 97}, 050401 (2006)

\bibitem{CEF2:2012} P. Calabrese, F. H. L. Essler, and M. Fagotti,
J. Stat. Mech. P07022 (2012).

\bibitem{CEF:2011}P. Calabrese, F.H.L. Essler, and M. Fagotti, Phys. Rev. Lett. {\bf 106}, 227203 (2011)

\bibitem{FCG:2011} 
L. Foini, L. F. Cugliandolo, and A. Gambassi, Phys. Rev. B {\bf 84},
212404 (2011);  arXiv:1207.1650.

\bibitem{sachdev} S. Sachdev, {\em Quantum phase transitions},
  Cambridge University Press. 

\bibitem{ising-ca}
J. Simon, W. S. Bakr, R. Ma, M. E. Tai, P. M. Preiss, and  M. Greiner,
Nature {\bf 472}, 307 (2011).

\bibitem{mc}
E. Barouch, B. McCoy, and M. Dresden, Phys. Rev. A {\bf 2}, 1075 (1970).

\bibitem{CEF1:2012} P. Calabrese, F. H. L. Essler, and M. Fagotti,
J. Stat. Mech. P07016 (2012).

\bibitem{SuppMat} See Supplemental Material at [\dots] for an outline of the form-factor approach.

\bibitem{FFs}
G. von Gehlen, N. Iorgov, S. Pakuliak, V. Shadura and Y. Tykhyy,
J. Phys. {\bf A 41}, 095003 (2008);
N. Iorgov, V. Shadura and Yu. Tykhyy, J. Stat. Mech. (2011) P02028.

\bibitem{IR:2011b}
H. Rieger and F. Igl\'oi, Phys. Rev. B. {\bf 84}, 165117 (2011). 



\bibitem{finiteT}
F. H. L. Essler and R. M. Konik, Phys. Rev. B {\bf 78}, 100403 (2008);
F. H. L. Essler and R. M. Konik, J. Stat. Mech. P09018 (2009).


\end{thebibliography}
\end{document}